\documentstyle[aaspp4,11pt]{article}

\def\gsim{\;\rlap{\lower 2.5pt
\hbox{$\sim$}}\raise 1.5pt\hbox{$>$}\;}
\def\lsim{\;\rlap{\lower 2.5pt
   \hbox{$\sim$}}\raise 1.5pt\hbox{$<$}\;}
\def\msun{{\rm\,M_\odot}}

\def\spose#1{\hbox to 0pt{#1\hss}}
\def\lta{\mathrel{\spose{\lower 3pt\hbox{$\mathchar''218$}}
     \raise 2.0pt\hbox{$\mathchar''13C$}}}
\def\gta{\mathrel{\spose{\lower 3pt\hbox{$\mathchar''218$}}
     \raise 2.0pt\hbox{$\mathchar''13E$}}}
\newcommand{\beq}{\begin{equation}}
\newcommand{\eeq}{\end{equation}}



\lefthead{MENOU, PERNA \& HERNQUIST} 
\righthead{Supernova Fallback Disks}

\begin{document}

\title{Stability and Evolution of Supernova Fallback Disks}

\author{Kristen Menou,\altaffilmark{1}}
 
\affil{Princeton University, Department of Astrophysical Sciences,
Princeton NJ 08544, USA, kristen@astro.princeton.edu}

\author{Rosalba Perna\altaffilmark{2} and Lars Hernquist}
 
\affil{Harvard-Smithsonian Center for Astrophysics, 60 Garden Street,
Cambridge MA 02138, USA, rperna@cfa.harvard.edu,
lars@cfa.harvard.edu}
 
\altaffiltext{1}{Chandra Fellow}
\altaffiltext{2}{Harvard Junior Fellow}

\begin{abstract}

We show that thin accretion disks made of Carbon or Oxygen are subject
to the same thermal ionization instability as Hydrogen and Helium
disks. We argue that the instability applies to disks of any metal
content. The relevance of the instability to supernova fallback disks
probably means that their power-law evolution breaks down when they
first become neutral. We construct simple analytical models for the
viscous evolution of fallback disks {to show that it is possible
for these disks to become} neutral when they are still young (ages of
a few $10^3$ to $10^4$~years), compact in size (a few $10^9$~cm to
$10^{11}$~cm) and generally accreting at sub-Eddington rates ($\dot M
\sim$ a few $10^{14}-10^{18}$~g~s$^{-1}$).  Based on recent results on
the nature of viscosity in the disks of close binaries, we argue that
this time may also correspond to the end of the disk activity
period. Indeed, in the absence of a significant source of viscosity in
the neutral phase, the entire disk will likely turn to dust and become
passive.  We discuss various applications of the evolutionary model,
including anomalous X-ray pulsars and young radio pulsars.  Our
analysis indicates that metal-rich fallback disks around newly-born
neutron stars and black holes become neutral generally inside the
tidal truncation radius (Roche limit) for planets, at $\approx
10^{11}$~cm. Consequently, the efficiency of the planetary formation
process in this context will mostly depend on the ability of the
resulting disk of rocks to spread via collisions beyond the Roche
limit.  It appears easier for the merger product of a doubly
degenerate binary, whether it is a massive white dwarf or a neutron
star, to harbor planets because its remnant disk has a rather large
initial angular momentum, which allows it to spread beyond the Roche
limit before becoming neutral. The early super-Eddington phase of
accretion is a source of uncertainty for the disk evolution models
presented here.

\end{abstract}

{\it subject headings}: X-ray: stars -- accretion, accretion disks --
supernovae: general -- pulsars: general -- stars: neutron -- MHD --
turbulence

\section{Introduction}

Fallback is a standard ingredient of contemporary core-collapse
supernova scenarios. It occurs when the reverse shock, triggered by
the impact of the ejected core material against the stellar envelope,
reaches the central compact object (Colgate 1988; Chevalier 1989; see,
e.g., Fryer \& Heger 2000 for recent core-collapse numerical
simulations). Some of the fallback material will have a specific
angular momentum in excess of the Keplerian value at the surface of
the central compact object and is therefore expected to settle
into a disk as it cools. The evolution of supernova fallback disks has
rarely been investigated.  Michel \& Dessler (1981, 1983) and Michel
(1988) first pointed out the role that such a disk could play for the
evolution of a young neutron star. Lin, Woosley \& Bodenheimer (1991)
considered the evolution of a fallback disk in more detail, with a
particular emphasis on the planetary formation process around young
neutron stars. Recently, Chatterjee, Hernquist \& Narayan (2000)
developed a model for Anomalous X-ray Pulsars (AXPs) based on the
viscous evolution of a fallback disk and its magnetospheric
interaction with the newly-formed neutron star. Alpar (1999, 2000)
also presents an accretion scenario for AXPs. Marsden, Lingenfelter \&
Rothschild (2001a; see also Marsden, Lingenfelter \& Rothschild 2001b)
have recently considered the effect of fallback disks on the spin
evolution of radio pulsars.

In this paper, we re-examine the evolution of supernova fallback disks
from a general perspective. The viscous evolution of fallback disks
should proceed according to the similarity solution derived by
Cannizzo, Lee \& Goodman (1990; see also Pringle 1974; Lynden-Bell \&
Pringle 1974) to study the accretion of the debris of a star that was
tidally disrupted by a supermassive black hole.  Numerical simulations
(Cannizzo et al. 1990) show that after an initial, transient phase,
the mass accretion rate drops off steadily as the supply of mass
dwindles. The long term evolution of the accretion rate is then well
approximated by a power-law decay with time.  During their evolution,
however, fallback disks enter a regime where they accrete at rates
comparable to those at which disks in close binaries become subject to
the thermal ionization instability ($\sim 10^{16}$~g~s$^{-1}$ or
so). The occurrence of the instability, which is at the origin of the
large amplitude outbursts of Dwarf Novae and X-ray Transients (see
reviews by Cannizzo 1993b; Osaki 1996; Lasota 2001), could signal the
end of the power-law evolution of fallback disks.

We investigate this possibility here, and explore its consequences for
the accretion scenario of AXPs and the planetary formation process
around young compact objects. {Meyer-Hofmeister (1992) and
Mineshige, Nomoto \& Shigeyama (1993) were the first to discuss the
role of the thermal ionization instability for supernova fallback
disks. Their work was limited to the case where hydrogen is
responsible for the instability and possibly leads to outbursts in the
disk. The contribution in this paper is new in that we extend the
relevance of the thermal ionization instability to metal-rich disks
and we focus on the possibility that the instability could instead
lead to the end of mass accretion in the disk. Note that, while
this work was sparked by the recent applications of the fallback
disk model described above, our results can be generally applied
to isolated metal-rich disks formed by any mechanism.}

The outline of the paper is as follows. In \S2, we show that metal-rich
disks, such as supernova fallback disks, are subject to the well-known
thermal ionization instability. In \S3, we determine the typical
characteristics of fallback disks at the time of the onset of the
instability and argue that this could also be the end of the
activity phase for the disk. In \S4, we discuss applications of the
scenario described in \S3 to AXPs, the planetary
formation process around newly-formed compact objects, doubly
degenerate mergers, and other interesting cases. We comment on some of
the limitations of our work in \S5 before concluding in \S6.

\section{Thermal Ionization Instability in Metal Disks}

\subsection{Method of Calculation}

The thermal-viscous stability of gaseous, thin accretion disks around
compact objects has previously been investigated in many different
contexts. The various existing results agree well with each
other and show that Hydrogen-dominated disks are thermally and
viscously unstable in regions of partial ionization corresponding to
disk central temperatures $T_c \sim 10^4$~K (or, equivalently, disk
effective temperatures $T_{\rm eff} \lsim 10^4$~K; Meyer \&
Meyer-Hofmeister 1981). Most of these studies considered disks of
solar-composition material (see, e.g., Cannizzo 1993a; Ludwig,
Meyer-Hofmeister \& Ritter 1994; Hameury et al. 1998 for recent
calculations), as expected in close binaries where the mass accreted
into the disk is supplied by a stellar companion. Smak (1983) was the
first to study the stability of disks made of pure Helium, while
Cannizzo (1984) presented a simplified time-dependent study of the
evolution of such a Helium disk (see also Tsugawa \& Osaki 1997;
El-Khoury \& Wickramasinghe 2000).

\begin{figure}
\plotone{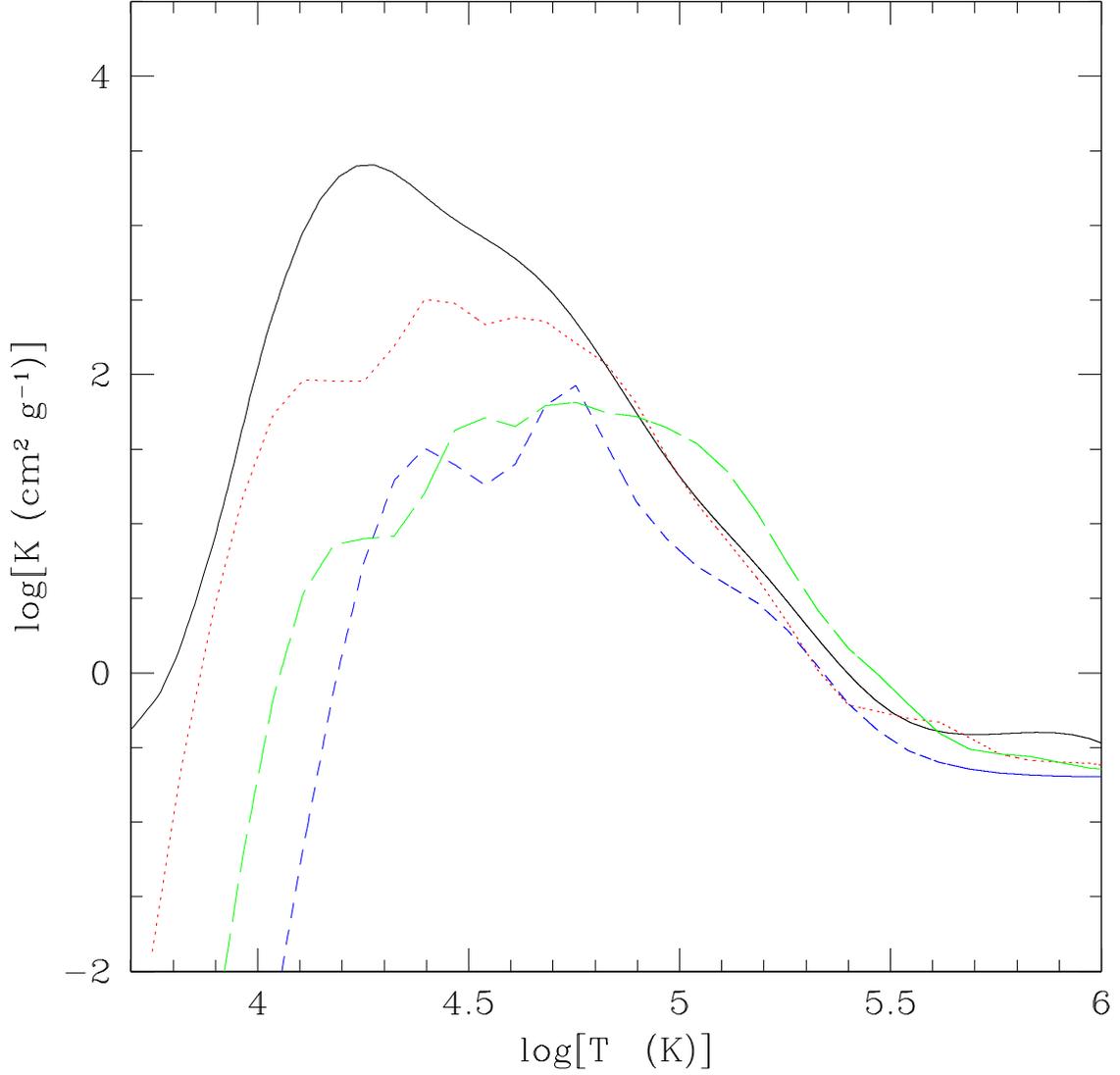}
\caption{Rosseland-mean opacities as a function of temperature, at a
mass density of $10^{-6}$~g~cm$^{-3}$, for solar composition
material (solid line), pure Helium (short-dashed), pure Carbon
(dotted) and pure Oxygen (long-dashed). In each case, the sudden
opacity drop at $T \lsim 10^4$~K corresponds to the recombination of
the last available free electron.
\label{fig:one}}
\end{figure}

We focus here on disks made of metals. Because of the limited
availability of pure metal opacities in the appropriate density and
temperature ranges, we restrict our work to pure Carbon and pure
Oxygen compositions. The case of solar composition material and pure
Helium are also considered for comparison with previous results. The
Rosseland-mean opacities were all taken from the OPAL database
(Iglesias \& Rogers 1996)\footnote{The OPAL database website is {\tt
http://www-phys.llnl.gov/Research/OPAL/}}. The opacities, at a
specific mass density of $10^{-6}$~g~cm$^{-3}$, are shown as a
function of temperature in Fig.~\ref{fig:one}, for the four
compositions of interest: solar composition (solid line), pure Helium
(short-dashed), pure Carbon (dotted) and pure Oxygen (long-dashed). In
each case, the sudden opacity drop at temperatures $T \lsim 10^4$~K
corresponds to the recombination of the last free electron. This
temperature differs from element to element because of different
ground state ionization potentials. The plateau in the opacity curves
at temperatures $T \gsim 10^5$~K corresponds to the electron
scattering limit. We are mostly interested in low-temperature
opacities for the present study.

The disk thermal equilibria are found by equating the local viscous
dissipation rate $Q^+$ to the radiative cooling rate $Q^-$, where
$Q^-$ is a function of the disk opacity. We neglect X-ray irradiation
when calculating the thermal equilibria, but we consider its effects
when discussing the evolution of supernova fallback disks in \S3. The
thermal equilibria are calculated from an extended grid of detailed
models for the disk vertical structure, for different values of the
surface density $\Sigma$ and the central temperature $T_c$, at a given
radius $R$ from the central object of mass $M_1$ (see Hameury et
al. 1998 for details on the numerical technique employed)\footnote{We
also checked the validity of our results by solving the simplified
vertically-averaged equation $Q^+ = (9/8) \nu \Sigma \Omega^2 = Q^-=4
\sigma T_c^4/(3 \Sigma \kappa)$, valid for an optically-thick disk,
where $\nu$ is the kinematic viscosity and $\kappa$ is the
opacity.}. Once a solution to the vertical structure (corresponding to
a given $\Sigma$, $T_c$ and $R$) is known, the disk effective
temperature $T_{\rm eff}$ is uniquely determined. The equation of
state adopted for the pure Helium, Carbon and Oxygen 
compositions is that of
a perfect gas with a monoatomic adiabatic index $\gamma=5/3$. A
detailed, tabulated equation of state is used for the solar
composition case (see Hameury et al. 1998 for details).

\subsection{Results}

\begin{figure}
\plotone{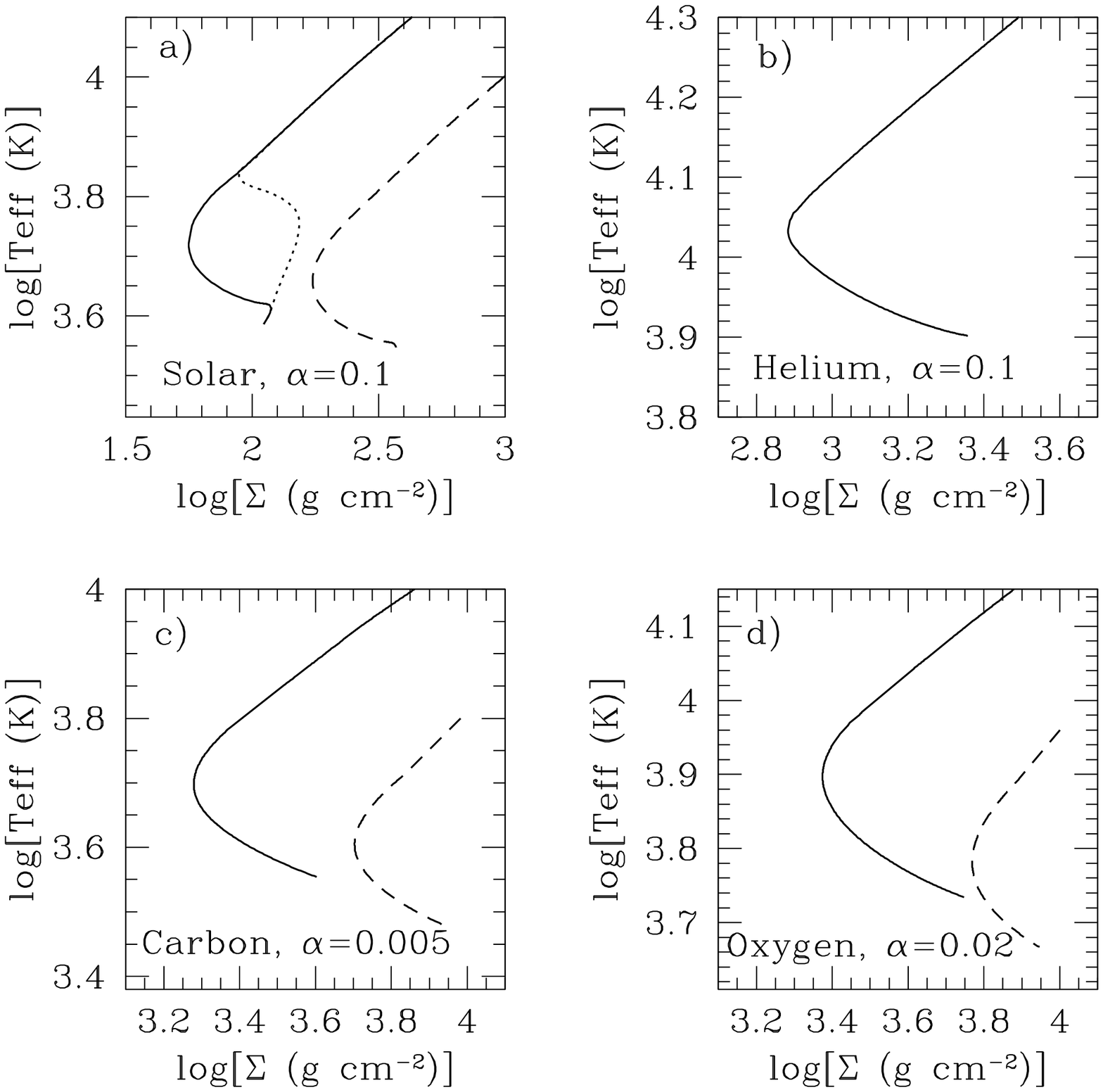}
\caption{Examples of thermal equilibrium curves (``S-curves'') for a
thin accretion disk, shown in a surface density vs. effective
temperature ($\Sigma - T_{\rm eff}$) diagram. The curves correspond to
a disk annulus located at $2 \times 10^{10}$~cm (solid line) or $6
\times 10^{10}$~cm (dashed line) from a central neutron star of mass
$M_1=1.4 M_\odot$. The four panels correspond to disks made of solar
composition material (a), pure Helium (b), pure Carbon (c) and pure
Oxygen (d). The value adopted for the viscosity parameter $\alpha$ is
indicated in each panel. In panel (a), the dotted line shows the
effect of allowing for convection in the disk vertical structure. In
each case, the regions of the curves which have negative slopes
indicate a thermally and viscously unstable disk (see text for
details).
\label{fig:two}}
\end{figure}

Figure~\ref{fig:two} shows examples of thermal equilibrium curves
(usually called ``S-curves'') in a surface density vs. effective
temperature diagram.  In each panel, the sections of the S-curves with
negative slope are thermally and viscously unstable (see, e.g.,
Cannizzo 1993b for a review of the instability). Note that the only
complete S-curves are those corresponding to the disk of solar
composition, because this is the only case for which opacities in the
neutral phase are available. In the solar composition case (having a
detailed equation of state), we also show the typical effect on the
S-curve when allowing for convection in the disk vertical structure
(dotted line), in regions where the vertical temperature gradients are
superadiabatic: the location of the unstable branch of the S-curve is
somewhat affected. Since the exact role and magnitude of convection in
thin accretion disks is not well understood, Fig.~\ref{fig:two}a
basically shows that convection is a source of uncertainty for the
derived stability criterion. The thermal equilibrium solutions are
shown for different values of the viscosity parameter $\alpha$ to
guarantee that opacities are available, for each composition, in the
range of density and temperature covered by the model. The conclusions
drawn below are essentially independent of the specific value of
$\alpha$ adopted.

The presence of regions of negative slope in the curves of thermal
equilibrium shown in Fig.~\ref{fig:two} indicates that the
corresponding disks are subject to the thermal ionization
instability. The well-known results for a solar composition and a pure
Helium disk are recovered here and the relevance of the instability is
extended to metal-rich disks. Although this was only shown for pure
Carbon and Oxygen compositions, we argue below that it is natural to
generalize this result to disks of any composition.

The instability occurs at a characteristic effective temperature
$T_{\rm eff, crit}$ (where the slope in the S-curve changes from
positive to negative), which is a function of the disk composition and
the radius of interest.\footnote{We checked that, whether the disk is
made of solar composition material, pure Helium, Carbon or Oxygen, the
value of $T_{\rm eff, crit}$ depends only weakly on the viscosity
parameter $\alpha$.}  The dashed and solid lines in Fig.~\ref{fig:two}
(corresponding to $R=2 \times 10^{10}$ and $6 \times 10^{10}$~cm,
respectively) show that the variation of the equilibrium curves with
radius is comparable for solar composition and metal-rich disks. On the
other hand, the systematic offset of $T_{\rm eff, crit}$ with
composition can be related to the different ground state ionization
potential in each case. Figure~2 of Mihalas et al. (1990), for
example, shows that, for the densities relevant here and in order of
decreasing temperature, Helium fully recombines first, followed by
Oxygen, and then Hydrogen and Carbon with comparable values. This
ordering is reflected in the temperatures at which the opacities drop
at recombination in Fig.~\ref{fig:one} and the ordering of $T_{\rm
eff, crit}$ for the various compositions shown in
Fig.~\ref{fig:two}.\footnote{Note that the relation between the disk
central temperature (which determines the ionization state) and its
effective temperature depends on the vertically integrated optical
thickness of the disk. The ordering of $T_{\rm eff, crit}$ therefore
also depends on the gas opacity in each case.} Other metals will have
different ground state ionization potentials but there is every reason
to expect that a disk of any composition will be subject to the same
ionization instability.

\subsection{Global Stability Criterion}

The curves of thermal equilibrium can be used to derive a global
stability criterion for the corresponding accretion disk. The relation
$\dot M \propto T_{\rm eff}^4$ between the disk accretion rate and its
effective temperature implies that a disk becomes thermally unstable if it
locally accretes at a rate below the value corresponding to $T_{\rm
eff, crit} (R)$.\footnote{The viscous instability, which follows the
exact same stability criterion, can generally be neglected because of
its much longer growth time.} The well-known stability criterion for a
solar composition disk is therefore (e.g. Hameury et al. 1998):
\begin{equation}
\dot M_{\rm crit}^+ \left( R \right) \simeq 9.5 \times 10^{15} m_1^{-0.9}
R_{10}^{2.68}~{\rm g~s}^{-1},  
\label{eq:ham98}
\end{equation}
where $m_1$ is the mass of the central object in solar units and
$R_{10}$ is the radius of interest in units of $10^{10}$~cm.  A disk
which satisfies everywhere $\dot M(R) > \dot M_{\rm crit}^+ (R)$ is
ionized and thermally (as well as viscously) stable. According to the
results presented above, the stability criterion for a disk made of
Helium or metals is quite similar, except for a different absolute
scaling due to the eventual offset in the value of $T_{\rm eff,
crit}$, as shown in Fig.~\ref{fig:two}. For instance, the value of
$T_{\rm eff, crit}$ for a Helium disk being approximately twice that
for a solar composition disk, the corresponding value of $\dot M_{\rm
crit}^+ (R)$ should be multiplied by a factor of $8$ or so. Since the
deviation in Eq.~(\ref{eq:ham98}) from the scaling $m_1^{-1} R^3$
followed by the local viscous dissipation rate (e.g. Frank, King \&
Raine 1992) arises from small variations with radius of the disk
density, and therefore opacity, at the temperature characteristic of
recombination, one may also expect slight deviations in the scaling of
$\dot M_{\rm crit}^+ $ with $m_1$ and $R$ for non Hydrogen-dominated
compositions.

\section{Power-Law Evolution of a Supernova Fallback Disk}

In this section, we apply the results derived in \S2 to the evolution
of a supernova fallback disk. One may expect the self-similar
evolution of this type of disk to break down when the thermal
instability first sets in. This is because the evolution of the disk
during the self-similar phase is regulated by viscosity, and when the
disk recombines, the nature and magnitude of the viscosity 
will likely change (see
\S3.3 for further discussion on this issue). In the following, we
estimate the time at which the thermal instability first sets in, and
what the corresponding disk parameters are.

\subsection{Non-irradiated Disk}

In order to simplify the analysis below, we adopt the following global
stability criterion to determine when a metal-rich disk becomes
neutral:
\begin{equation}
\dot M_{crit}^+ \left( R \right) \approx \beta 10^{16} R_{10}^{3}~{\rm
g~s}^{-1}.
\label{eq:crit}
\end{equation}
This expression differs only slightly from that given in
Eq.~(\ref{eq:ham98}). We also neglect the very weak dependence on the
viscosity parameter $\alpha$. Given that we ignore the exact
composition of the disk (which determines the exact value of $T_{\rm
eff,~crit}$ at which the disk becomes neutral), that we neglect the
effects of convection on the disk vertical structure and that we are
interested below in radii which are in the vicinity of $10^{10}$~cm,
the use of equation~(\ref{eq:crit}) is justified at an
order-of-magnitude precision level or so. The factor $\beta$ in
Eq.~(\ref{eq:crit}) captures all these uncertainties and any
dependence on $m_1 \neq 1$ (compare Eq.~[\ref{eq:ham98}]).

In a steady (or quasi-steady) disk, the criterion is first met at the
disk outer edge (where the gas in the disk becomes neutral first).
Assuming that the disk evolves according to the self-similar solution
of Cannizzo et al. (1990; for an electron scattering
opacity),\footnote{Cannizzo et al. found that the power-law evolution
for a standard (Hydrogen-dominated) Kramers opacity is very similar to
that for an electron scattering opacity. We use their analytical
solution for the latter case here, which should be reasonably accurate
even for metal-rich disks.} after an initial transient phase of
duration $t_0$, the disk mass, $M_d$, outer radius, $R_d$, and mass
accretion rate, $\dot M_d$, obey the relations:
\begin{eqnarray}
M_d (t)  & = & M_d (t_0) \left( \frac{t}{t_0} \right)^{-3/16} \label{eq:evol1},\\
R_d (t)  & = & R_d (t_0) \left( \frac{t}{t_0} \right)^{3/8} \label{eq:evol2},\\
\dot M_d (t)  & = & \dot M_d (t_0) \left( \frac{t}{t_0} \right)^{-19/16}.
\label{eq:ssevol} \label{eq:evol3}
\end{eqnarray}
The equation for the disk radius results from the assumptions that
most of the disk mass resides in its outermost regions and that the
total angular momentum of the disk is conserved. One can easily check
that most of the disk mass is indeed located at the outer edge in the
power law solutions used here, and that little angular momentum is
accreted with the gas that reaches the central compact object. The
disk radius calculated in this way should approximate reasonably well the
actual disk radius, in the limit of long time evolution and a large
ratio of outer to inner disk radii (see Fig.~2 of Cannizzo et
al. 1990 for a comparison of the results between a similarity solution
and a detailed numerical integration).

The quantity 
\begin{equation}
\frac{\dot M_d (t)}{R_d^3 (t)}=\frac{\dot M_d (t_0)}{R_d^3 (t_0)} \left(
\frac{t}{t_0}\right)^{-37/16}
\end{equation}
is a measure of the amount of viscous dissipation at the disk outer
edge, as a function of time. According to Eq.~(\ref{eq:crit}), when
\begin{equation}
\frac{\dot M_d (t)}{R_{d,10}^3 (t)} \approx \beta 10^{16}~{\rm
g~s}^{-1},
\label{eq:dead}
\end{equation}
the outermost annulus of the disk becomes neutral.

Let $M_d(t_0)$ and $R_d(t_0)$ be the initial mass and radius of the
fallback material ending up in the disk. Various arguments concerning
supernova explosions suggest that the amount of fallback is $< 0.1
\msun$ in the case of the formation of a neutron star (Lin, Woosley \&
Bodenheimer 1991; Chevalier 1989), while conservation of angular
momentum during the supernova explosion and the subsequent fallback
would guarantee that the material brought back to the compact object
by the reverse shock ends up at radii $\lsim 10^9$~cm because of its
initial location within the stellar core (of typical size a white
dwarf radius). Recent numerical simulations of the pre-supernova
evolution of rotating stars suggest values $j \approx
10^{16}-10^{17}$~cm$^2$~s$^{-1}$ for the specific angular momentum of
the material surrounding the iron core just before collapse (Heger,
Langer \& Woosley 2000). This corresponds to Keplerian radii $\approx
10^6-10^8$~cm around a solar mass central object. Below, when we
specify the initial radius of a fallback disk, $R_d(t_0)$, it
corresponds to choosing the mean specific angular momentum of the
material forming the disk, or equivalently to choosing the total disk
angular momentum $\propto M_d(t_0) R_d^{1/2}(t_0)$ for a given initial
disk mass $M_d(t_0)$.

The fallback material with excess angular momentum, after cooling,
will spread via a disk on a typical local viscous timescale, that we
identify with the duration of the initial transient accretion phase
(see Figure~3b of Cannizzo et al. 1990):
\begin{equation}
t_0 \equiv \frac{R^2 \Omega_K}{\alpha c_S^2} \approx 6.6 \times
10^{-5} (T_{c,6})^{-1} R_{d,8}^{1/2}(t_0)~{\rm yrs},
\label{eq:vistime}
\end{equation}
where $\Omega_K$ is the local Keplerian angular speed, $c_S$ is the
sound speed, $R_{d,8}(t_0)$ is the initial radius in units of
$10^8$~cm and $T_{c,6}$ is a typical temperature in the disk during
this early phase, in units of $10^6$~K. A viscosity parameter
$\alpha=0.1$ and a solar mass central object were assumed when
deriving Eq.~(\ref{eq:vistime}). Note that $10^6$~K is a
representative temperature for a disk annulus located at a radius
$\sim 10^8$~cm from a stellar-mass central object accreting at the
Eddington rate (see, e.g., Fig.~5 in Hameury et al. 1998 for typical
temperatures).\footnote{{Note that at super-Eddington accretion
rates, the flow could in principle be much hotter than assumed here,
reducing the initial viscous timescale ($t_0$)
accordingly. Uncertainties related to the super-Eddington accretion
phase are further discussed in \S5.}} This temperature explicitly
depends on the radius of interest as well (in a non-trivial way),
which is why we keep the scaling with $T_{c,6}$ explicit in the above
equation for $t_0$.

The accretion rate at the onset of the self-similar evolution phase is
$\dot M_d (t_0) \sim M_d (t_0)/ t_0$ (i.e. after the transient phase;
see again Fig.~3b of Cannizzo et al. 1990), so that by combining
Eqs.~(\ref{eq:evol1}-\ref{eq:evol3}) and~(\ref{eq:dead}), we find that
the outermost disk annulus becomes neutral after a time:
\begin{equation}
t_n \approx 1.5 \times 10^3~{\rm yrs} ~\left( \frac{M_d (t_0)}{\beta
10^{-3}~\msun} \right)^{16/37} T_{c,6}^{-21/37} R_{d,8}^{-75/74}
(t_0).
\end{equation}
This happens for a disk outer radius
\begin{equation}
R_{d,8} (t_n) \approx 570 ~\left( \frac{M_d (t_0)}{\beta
10^{-3}~\msun} \right)^{6/37} T_{c,6}^{6/37} R_{d,8}^{16/37}(t_0),
\end{equation} 
a disk mass
\begin{equation}
M_d (t_n) \approx 4.2 \times 10^{-5}~\msun ~\beta^{3/37}~\left( \frac{M_d
(t_0)}{10^{-3}~\msun} \right)^{34/37} T_{c,6}^{-3/37}
R_{d,8}^{21/74} (t_0),
\end{equation} 
and a disk accretion rate
\begin{equation}
\dot M_d (t_n) \approx 1.85 \times 10^{18}~{\rm g~s}^{-1} ~\beta^
{19/37}~\left( \frac{M_d (t_0)}{10^{-3}~\msun} \right)^{18/37}
T_{c,6}^{18/37} R_{d,8}^{48/37} (t_0).
\end{equation}

For a low mass and low angular momentum disk ($M_d (t_0)=10^{-6}~\msun$,
$R_{d,8}(t_0) = 0.01$ and $\beta =1$), this gives:
\begin{eqnarray}
t_n & \approx& 7.5 \times 10^3~{\rm yrs}~(\times T_{c,6}^{-21/37}), \\ 
R_d (t_n) & \approx& 2.5 \times 10^{9}~{\rm cm}~(\times T_{c,6}^{6/37}),\\ 
\dot M_d (t_n) &\approx& 10^{14}~{\rm g~s}^{-1}~(\times T_{c,6}^{18/37}).
\end{eqnarray}

For a high mass and high angular momentum disk ($M_d (t_0)=10^{-2}~\msun$,
$R_{d,8}(t_0) = 1$ and $\beta =1$), this gives:
\begin{eqnarray}
t_n & \approx& 4 \times 10^3~{\rm yrs}~(\times T_{c,6}^{-21/37}), \\ 
R_d (t_n) & \approx& 8.3 \times 10^{10}~{\rm cm}~(\times T_{c,6}^{6/37}),\\ 
\dot M_d (t_n) &\approx& 5.7 \times 10^{18}~{\rm g~s}^{-1}~(\times T_{c,6}^{18/37}).
\end{eqnarray}
A typical disk surface density is then:
\begin{equation}
\Sigma \sim \frac{M_d (t_n)}{R_{d}^2 (t_n)} \gg 1~{\rm g~cm}^{-2},
\end{equation}
so that the disk is still very optically thick to X-rays and its own
radiation.

\subsection{Irradiated Disk}

The self-similar solutions to the viscous evolution of a thin
accretion disk derived by Cannizzo et al. (1990) were obtained under
the assumption of purely local viscous heating for the disk. This is
clearly not appropriate for an irradiated disk, which is subject to an
additional non-local heating. One can easily show, however, that even
in the strong and steady irradiation limit (when the irradiation flux
is time-independent and dominates over local viscous heating), the
viscous evolution of the disk still obeys a similarity solution, with
larger power law indices.  Vrtilek et al. (1990; see also Cunningham
1976) showed that a vertically isothermal disk (satisfying hydrostatic
equilibrium) has a vertical height $H$ which varies as a function of
radius as $R^{9/7}$. The viscosity in this strongly irradiated
$\alpha$-disk is then $\nu \equiv \alpha H^2 \Omega_K \propto
R^{15/14}$ (compare to $\nu \propto \Sigma^{2/3} R$ for the
non-irradiated case; Cannizzo et al. 1990). Lynden-Bell \& Pringle
(1974) have solved the general problem of the viscous evolution of a
disk with kinematic viscosity $\nu \propto R^n$, for an arbitrary
value of $n$. Using their similarity solutions (valid for $n \neq 2$),
we find that a strongly and steadily irradiated $\alpha$-disk evolves
viscously in a self-similar manner with a power law index $l=1/(4-2n)$
for the time evolution of the disk mass, so that $M_d(t) \propto
t^{-7/13}$ and $\dot M_d(t) \propto t^{-20/13}$. This corresponds to a
more rapid evolution than in the non-irradiated solution described by
Eqs.~(\ref{eq:evol1}-\ref{eq:evol3}). The strong-irradiation solution
largely overestimates the strength of irradiation in that it assumes a
constant flux hitting the disk. In a more realistic scenario, the
viscosity $\nu$ will also be a decreasing function of time, as the
accretion rate and the irradiation flux decrease with time.

The ionization stability criterion for an irradiated disk is also
different because irradiation, as an extra source of heating, is able
to keep the disk ionized where it would have been neutral
otherwise. The stability of irradiated disks has previously been
investigated in the context of accretion in low-mass X-ray binaries
(see, e.g., Van Paradijs 1996; Dubus et al. 1999 and references
therein).  The intrinsic emission from the hot, young neutron star can
be neglected for the evolution of a supernova fallback disk because it
does not contribute more than
$10^{33}$~ergs~s$^{-1}$.\footnote{Similarly, we neglect the extra
source of heating due to radioactive decay because it only concerns
the very early evolution of the supernova fallback material (see,
e.g., Michel 1988).} To account for the effect of accretion-induced
irradiation, we adopt the following global stability criterion:
\begin{equation}
\dot M_{crit}^{irr} \left( R \right) \approx \beta_{irr} 10^{15}
R_{10}^{2}~{\rm g~s}^{-1}.
\label{eq:critirr}
\end{equation}
This expression differs only slightly from the formula given by Dubus
et al. (1999): $\dot M_{crit}^{irr} (R) \simeq 1.5 \times 10^{15}
m_1^{-0.4} R_{10}^{2.1}~{\rm g~s}^{-1}$. Again, the use of
Eq.~(\ref{eq:critirr}) simplifies the derivation below and is
justified at an order-of-magnitude precision level or so.  A specific
calibration for the disk geometry and albedo in low mass X-ray
binaries (see Dubus et al. 1999) was used to derive the absolute
scaling in Eq.~(\ref{eq:critirr}).  Consequently, the factor
$\beta_{irr}$ in Eq.~(\ref{eq:critirr}) covers uncertainties
concerning the disk composition, geometry and albedo, as well as any
dependence on $m_1 \neq 1$.

A comparison between Eq.~(\ref{eq:crit}) and Eq.~(\ref{eq:critirr})
suggests that disk irradiation dominates the stability properties of
disks which extend beyond $\sim 10^9$~cm.  The solutions derived in
\S3.1 are therefore valid only for disk initial parameters such that
the disk becomes neutral before it reaches radii $\sim 10^9$~cm or so.
If irradiation becomes important during the disk evolution, the
quantity $\dot M_d / R_d^2$, which measures the local heating due to
disk irradiation, must this time be used to determine the disk
stability properties.  As a fallback disk spreads, it may reach radii
$\sim 10^9$~cm, where irradiation starts becoming important for its
energy budget. The disk evolution accelerates slowly as the disk
spreads further and irradiation becomes more and more important
(driving the disk more and more toward the isothermal case). The exact
solution to this problem can only be obtained numerically (especially
because the viscosity is then a function of time).

In Appendix~A, we consider two simple cases for the evolution of an
irradiated disk.  In one case, labeled ``weakly-irradiated disk,'' we
assume that the power-law evolution of the disk remains as slow as in
the non-irradiated case, while we determine the stability properties
of the disk including the effect of irradiation
(Eq.~[\ref{eq:critirr}]). In the other case, labeled
``strongly-irradiated disk,'' we assume that as soon as the disk
reaches $10^9$~cm and irradiation becomes significant, its evolution
becomes more rapid and follows the similarity solution derived above
for a vertically isothermal disk. The main result obtained from these
models is that, even when the effects of irradiation are taken into
account, a fallback disk is expected to become neutral when it is
still young, compact and accreting at substantially sub-Eddington
rates (see Appendix~A). We emphasize that the ``weakly-irradiated''
case is probably a better approximation than the
``strongly-irradiated'' one because of the extreme assumptions made
concerning the strength of irradiation in the latter case (Dubus et
al. 1999, for instance, note that even the irradiated disks in low
mass X-ray binaries are far from being vertically isothermal). In
Appendix~B, we also show that a relatively compact disk is a general
feature of the models that remains valid even for an arbitrarily fast
power law evolution of the fallback disk.

\subsection{Subsequent Evolution}

The immediate outcome of the thermal ionization instability, when it
is first triggered in the outermost regions of the ionized disks of
close binaries, has been studied numerically in detail and is
relatively well understood. It generally leads to the outside-in
propagation of a cooling (or equivalently recombination) front that
reaches the central compact object relatively quickly because the
instability is successively triggered in disk annuli at smaller and
smaller radii. The front propagation typically takes only days to tens
of days, depending on the role of irradiation at controlling its
propagation (see, e.g., Cannizzo 1993a; Menou, Hameury \& Stehle 1999;
Dubus, Hameury \& Lasota 2001). Because of the strong analogy between
fallback disks and disks in X-ray Transients (in terms of size and
susceptibility to the same ionization instability), there is every
reason to believe that this phase should proceed identically in a
fallback disk. Right after the front propagation, the surface density
profile in the disk is approximately $\Sigma \propto R$, because of
the important redistribution of mass and angular momentum driven by
the front during its propagation (putting most of the mass in the
outer disk; see, e.g., Cannizzo 1993a; Hameury et al. 1998; Dubus et
al. 2001 for numerical examples).

The subsequent disk evolution, after it becomes entirely neutral, is
less certain. It mostly depends on the nature and magnitude of
viscosity in a neutral disk, which is not well understood. There are
many reasons to believe that the MHD turbulence resulting from the
magneto-rotational instability (Balbus \& Hawley 1991; 1998; Hawley,
Gammie \& Balbus 1996) cannot be self-sustained in the neutral disks
of X-ray Transients and Dwarf Novae (i.e. during quiescence) because
they are too weakly ionized (Gammie \& Menou 1998; Fleming, Stone \&
Hawley 2000; Menou 2000).\footnote{Note that Hall effects, which are
small in the present context but have a potentially destabilizing
effect, have not been included in these studies (Balbus \& Terquem
2001).} Convection does not provide the necessary outward angular
momentum transport for accretion to proceed (Ryu \& Goodman 1992;
Stone \& Balbus 1996; Cabot 1996), while the possibility of
hydrodynamical turbulence is not supported by the same simulations of
Keplerian disks which show the development of the Balbus-Hawley
instability in the magnetized case (Hawley, Balbus \& Winters
1999). Purely theoretical arguments have also been presented against
hydrodynamical turbulence (Balbus \& Hawley 1998; Hawley, Balbus \&
Winters 1999; but see Richard \& Zahn 1999 for a different view). This
leads to the interesting possibility that angular momentum transport
could be associated with the tidal influence of the companion star in
the neutral disks of close binaries (Spruit 1987), in which case
isolated disks like the fallback disks considered here could be
essentially passive when they become neutral (Menou
2000).\footnote{Another mechanism that has been proposed to transport
angular momentum in disks, the tidally-induced parametric instability,
also relies on the presence of a companion star (see Balbus \& Hawley
1998; Goodman 1993).}

We note that the results concerning MHD turbulence in neutral disks
established for Dwarf Novae and X-ray Transients cannot be directly
applied to fallback disks because of their different
composition.\footnote{In their study of disks around supermassive
black holes, Menou \& Quataert (2001) pointed out that the disk size
also matters for the quality of gas to magnetic coupling in the
neutral phase resulting from the thermal ionization instability. In
that respect, fallback disks are comparable to disks in dwarf novae
and X-ray transients.}  Indeed, the presence of MHD turbulence mainly
depends on the ionization fraction in the disk, which is essentially
an exponential function of the disk central temperature (via the Saha
equation in LTE; Gammie \& Menou 1998). The central temperature itself
depends on the opacity in the neutral/molecular phase ($T_c \propto
\tau^{1/4}$ in an optically thick disk at a given $T_{\rm eff}$; Frank
et al. 1992), so that the properties of magnetic coupling of the gas
in a metal disk could potentially be different from those in a solar
composition disk.

The opacity of a neutral metal-rich disk will be determined by a
complex chemistry (such as, for instance, the formation of CO
molecules in a metal-rich gas) which depends on the actual composition
and is presently difficult to characterize.  We note, however, that
Lin et al. (1991) advocate a mixture of Fe, Si, O, He and perhaps
traces of H in their discussion of supernova fallback disks (around a
neutron star). This is not a mixture that would allow the formation of
all the molecules whose contribution to the opacity is important in a
solar composition gas (in particular those based on carbon and
hydrogen chemistry; e.g. Alexander \& Ferguson 1994). This shows that
smaller opacities are actually quite possible in a fallback disk (as
compared to a solar composition disk) in the neutral/molecular phase.
If the fallback disk cools even further, the dust grains which start
forming would contribute much more to the disk opacity (per unit mass)
than in a solar composition gas because the disk is made entirely of
metals. But this time also corresponds to the end of the gaseous phase
for the disk as it presumably turns entirely to dust (contrary to a
solar composition disk) and becomes unlikely to sustain MHD
turbulence. Consequently, if the neutral/molecular opacities in
fallback disks are indeed not substantially larger than in the solar
composition case, a passive gaseous disk, soon to turn to dust,
appears as the plausible outcome of the evolution outlined in \S3.1
and \S3.2, given our current understanding of the nature of viscosity
in the disks of transient close binaries.  The time the disk first
becomes neutral will then signal the end of its activity period.

We adopt this view in what follows, although we comment on the
consequences of an eventual residual viscosity for neutral fallback
disks in \S5. We note that, even for passive neutral disks, there
could in principle still be some residual disk accretion in surface
layers which are ionized by the X-rays coming from the central, hot
neutron star (in a manner similar to that envisioned for T-Tauri disks
by Gammie 1996), if the disk geometry allows it. This possibility will
probably not exist in the case of accretion onto a central black hole
(weaker self-irradiation).

\section{Applications}

\subsection{Anomalous X-ray Pulsars}

\subsubsection{General Results}

The evolutionary scenario described in \S3 has a direct application as 
an accretion model for Anomalous X-ray Pulsars (AXPs), superseding that
developed by Chatterjee et al. (2000).  In this model, the evolution
of a newly-born neutron star mainly depends on the nature of its
magnetospheric interaction with the surrounding fallback
disk. Chatterjee et al.  find that, depending on the initial values of
the neutron star spin period, magnetic field strength and disk mass, a
system can evolve into an X-ray luminous AXP, where the neutron star
has been substantially slowed down by the torque exerted by the disk
during a long propeller phase, or into a radio pulsar which did not
interact as strongly with its surrounding disk (located beyond the
pulsar light cylinder at late times). Chatterjee \& Hernquist (2000)
showed that this scenario can reproduce the inferred distribution of
AXP spin periods, luminosities and age distributions for reasonable
and broad distributions of the model input parameters.

One of the shortcomings of the Chatterjee et al. model for
AXPs is the necessity to postulate that the end of the activity period
for the fallback disk (required to explain the inferred ages of AXPs)
corresponds to a transition to a radiatively inefficient
Advection-Dominated Accretion Flow (ADAF; see Narayan, Mahadevan \&
Quataert 1998 for a review). The AXP model also requires the onset of
a propeller effect in this late evolutionary phase to effectively
reduce the accretion rate onto the central neutron star.

These requirements are no longer needed if the viscous evolution
scenario described in \S3 is adopted.  In our model, the disk
always remains geometrically thin, and the
end of the disk activity
period (and therefore of the AXP activity) simply corresponds to the
time when the disk suddenly becomes neutral and passive under the
action of the thermal ionization instability. This model, using disk
accretion physics tested in close binary systems and no fine tuning of
parameters (except a choice of initial disk mass and angular momentum),
can provide a satisfying explanation for the inferred ages and
luminosities of AXPs. At the same time, this fallback disk scenario can
accommodate the constraints on disk sizes around AXPs because the end of
the activity period corresponds to such an early time, that the disk
did not have time to spread viscously very much. We have shown that,
if the fallback material ending up in the disk has low initial mass
and angular momentum, the model can account for disk sizes as small as
a few $ 10^{9}$~cm.  For example, a specific non-irradiated fallback
disk with $M_d (t_0)=10^{-4}~\msun$, $R_{d,8}(t_0) = 0.03$ and $\beta
=1$ becomes neutral when:
\begin{eqnarray}
t_n & \approx& 2 \times 10^4~{\rm yrs}~(\times T_{c,6}^{-21/37}), \\ 
R_d (t_n) & \approx& 8.6 \times 10^{9}~{\rm cm}~(\times T_{c,6}^{6/37}),\\ 
\dot M_d (t_n) &\approx& 6.4 \times 10^{15}~{\rm g~s}^{-1}~(\times T_{c,6}^{18/37}),
\end{eqnarray}
in good agreement with the inferred properties for AXPs.  We further
discuss the issue of disk size in \S4.1.2, together with that of the
strength of disk irradiation in AXPs. By virtue of the power law
evolution in time, the model predicts that most systems should be
observed relatively close to the end of their activity period.  The
fact that the systems are necessarily observed before $t_n$,
i.e. before the disk has reached its final radial extent, only helps
to explain the compact sizes inferred. In our view, the simplicity of
this model, the classic nature of its ingredients, and its ability to
reproduce reasonably well the observed characteristics of AXPs without
parameter fine-tuning make it a strong alternative to the
magnetar model (Thompson \& Duncan 1996; Heyl \& Hernquist 1997).

We note a possible difficulty for the proposed AXP model to explain
disk sizes much smaller than inferred to date. The magnetospheric
truncation radius of the disk is $R_m \approx 2.5 \times 10^8~{\rm
cm}~ B_{12}^{4/7} \dot M_{16}^{-2/7}$, where $B_{12}$ is the neutron
star surface magnetic field strength in units of $10^{12}$~G and $\dot
M_{16}$ the accretion rate is units of $10^{16}$~g~s$^{-1}$
(e.g. Frank et al. 1992).  Clearly, the disk outer radius cannot be
much smaller than $R_m$. We also note that the disk must acquire some
angular momentum as it spins down the neutron star. This gain is
moderate in the quasi-equilibrium ``tracking-phase'' model of
Chatterjee et al. (2000; see discussion in \S5) but its effect is to
increase the disk size compared to the predictions made in \S3. These
complications show that a detailed numerical integration is probably
required to obtain more reliable estimates of the disk characteristics
when it becomes neutral, especially in those cases where the initial
disk mass and angular momentum are small enough that the disk outer
radius, $R_{\rm out}$, is not much larger than the inner edge, $R_{\rm
in} = R_m$.

\subsubsection{Disk Size and Irradiation}

Upper limits to the optical flux from AXPs provide useful constraints
on all accretion models of these systems (see, e.g., Hulleman et
al. 2000a, Perna, Hernquist \& Narayan 2000; Perna \& Hernquist 2000
and references therein). The tightest existing constraint is that
given by the recent identification of a possible optical counterpart 
to the
AXP 4U~0142+61 (Hulleman, van Kerkwijk \& Kulkarni 2000b).  Hulleman
et al. (2000b) infer that only a compact accretion disk, of outer
radius $R_{\rm out} \lsim 0.05 R_\odot \approx 3.5 \times 10^9$~cm
(with uncertainties in the source distance and the strength of
irradiation), is allowed in 4U~0142+61. By analogy, one can expect
comparable constraints on the maximal radial extent of hypothetical
disks around other AXPs.

The evolutionary model for supernova fallback disks described in \S3
appears consistent with these restrictions: active fallback disks
should be compact because they do not have time to spread extensively
before they become neutral (and passive), when the thermal ionization
instability first sets in. We have shown that, with a choice of rather
small initial mass and angular momentum for the disk, a disk size of a
few $\sim 10^{9}$~cm may be expected at the end of the activity
phase.\footnote{{The disk sizes considered here are smaller than
in Hulleman et al. (2000a), Perna, Hernquist \& Narayan (2000), and
Perna \& Hernquist (2000). We attribute this difference partly to
smaller disk mass and angular momentum and partly to a different
description of the initial super-Eddington accretion phase. While
Perna et al. use the self-similar solution of Cannizzo et al. (1990)
in this regime for which it does not strictly apply, we assume here
that the early evolution occurs on a relatively long viscous timescale
associated with efficient cooling of the disk. None of these two
possibilities is fully satisfying but there is presently no robust
solution to this problem (see \S5).}}  Although this value is only an
estimate of the disk radius and the models are subject to various
uncertainties (see \S5 for a discussion), the possibility of
explaining very compact disks with a reasonable choice of initial disk
parameters is quite satisfying given the level of simplicity of the
model.

>From the optical constraints available so far, and the very high ratio
$F_X/F_{\rm opt} \gsim 10^4$ (Hulleman et al. 2000b) measured for the
AXP 4U 0142+61, it is clear that, if there is a disk in this source,
it must be of much smaller size and much less irradiated than typical
disks in low mass X-ray binaries.  The weakness of irradiation simply
results from the small disk size. For instance, at a typical AXP
luminosity of a few $10^{35}$~erg~s$^{-1}$, Hulleman et al.  (2000a)
show that the R band flux is typically dominated by viscous
dissipation for disk radii $\lsim$ a few $10^9$ to $10^{10}$~cm.  The
weak irradiation also means that the evolution model for a
non-irradiated disk described in \S3.1 is more appropriate than the
models with irradiation described in Appendix~A.

It is also worth mentioning that a careful comparison between the
light expected from a supernova fallback disk and the observational
constraints probably requires a model more elaborate than a simple
Shakura-Sunyaev $\alpha-$disk. Although it is true that the emission
from a disk evolving viscously as considered in \S3 is similar to that
of an $\alpha-$disk for the bulk of the disk, it is no longer true
close to the disk outer edge. There, the disk surface density, and the
viscous dissipation proportional to it, fall with radius faster than
in an $\alpha-$disk (see, e.g., Eq.~[13] in Cannizzo et
al. 1990). Given that most of the disk optical light (e.g. R band;
Hulleman et al. 2000a) comes from the outer regions for the typical
disk sizes considered, inferring an accurate limit on the disk radius
from the observations probably requires that this effect be taken into
account. We acknowledge that the disk radii as estimated in this paper
do not take this effect into account either.

\subsubsection{Observational Signatures}

If supernova fallback disks power AXPs, these systems may reveal
signatures of the presence of a disk (most likely in optical). The
best analogy would then be to disks in low mass X-ray binaries if
fallback disks are strongly irradiated, but the likelihood that they
are not significantly irradiated because of their small sizes make
disks in Cataclysmic Variables (CVs; see Warner 1995 for a review)
perhaps the best equivalent. A characteristic signature of these disk
is a broad, double-peaked emission line (function of orientation),
although the line may not be that broad if $R_{\rm in}=R_m$ is not
much smaller than $R_{\rm out}$ for fallback disks (see the analogy
with intermediate polars in Warner 1995).  Many of the atomic
transitions available to disks of solar composition may not be
available to fallback disks, however, because of their high
metallicity. It is still possible that (presumably weak) hydrogen
lines could be observed if the disk contains a small fraction of
hydrogen. In X-rays, the possibly high iron content of the (cold) disk
may favor energy emission through the iron K$\alpha$ fluorescence line (see,
e.g., George \& Fabian 1991), although a sufficient flux of hard
X-rays reprocessed by the disk may not be available to power the line
if supernova fallback disks are not significantly irradiated.

The variability properties of fallback disks may be quite different
from those of disks in CVs as well. The two variability phenomena
whose origin has been identified in CVs (flickering and dips) do not
apply to fallback disks because they do not possess a bright spot,
where the stream of mass transferred from the companion impacts the
disk in close binaries. Warner (1995) notes, however, that part of the
flickering seen in CVs is apparently intrinsic to the disk, so that
some flickering from supernova fallback disks may be expected.

If the optical light from AXPs were dominated by X-ray reprocessing,
it would follow any X-ray fluctuation, after a delay corresponding to
the light travel time over the disk radial extent. Optical pulsations
could potentially be observed despite the strong isotropization of the
X-ray emission due to general relativistic effects in the vicinity of
the neutron star (Perna \& Hernquist 2000).  In the case of the AXPs,
however, because of the very small size inferred for the disk, the
optical emission is likely dominated by viscous dissipation, so that
no optical pulsations would be expected.  In the magnetar
interpretation for the AXPs, no detailed predictions exist for the
mechanism of optical emission and its characteristics.  However, if
this emission is due to magnetospheric effects as in isolated pulsars
then, most likely, it will be highly pulsed as observed, for example,
in the case of the Crab pulsar (Warner et al. 1969).

\subsection{Radio Pulsars}

The model of Chatterjee et al. (2000) for the magnetospheric
interaction of a young neutron star with a surrounding fallback disk
predicts that many systems will not be X-ray luminous sources but
instead become standard radio pulsars. This happens when the disk
magnetospheric truncation occurs beyond the light cylinder in the
system. Chatterjee \& Hernquist (2000) estimate that, for reasonable
distributions of input model parameters, $\sim 80\%$ of newly born
neutron stars will follow the radio pulsar evolutionary path.

The evolution of a fallback disk in this context will first depend on
its capacity to survive in the environment of an
energetic pulsar wind. If
the disk survives, then it seems plausible that it will also evolve
according to the scenario described in \S3. The structure of the
magnetic field outside the light cylinder, $R_{\rm lc}$, is quite
uncertain, but it is clear that if the disk penetrates inside $R_{\rm
lc}$, it will encounter a strong centrifugal barrier there because the
magnetosphere rotates at a much faster rate than the local Keplerian
value. Ejection via the propeller effect would probably follow
(Illarionov \& Sunyaev 1975), allowing the disk to lose mass at its
inner edge (as required for the model in \S3 to apply).

If so, the neutron star will also be spun down by the propeller
effect. Determining the total amount of spin-down applied during the
disk activity period requires a detailed accretion+torque model that
will not be investigated here.  We note, however, that in the
magnetospheric scenario of Chatterjee et al. (2000), radio pulsars are
the results of the evolution of the young neutron stars with the least
massive disks. Over their activity period, which is of shorter
duration than for more massive disks, these disks will therefore apply
comparatively lower integrated spin down torques.  Order-of-magnitude
estimates suggest that the integrated torque applied by a low mass and
low angular momentum disk such as those considered in \S3 would
result in a period increase of at most a factor of a few for a radio
pulsar initially spinning at 15~ms. It appears therefore possible to
keep fast spinning radio pulsars in this scenario, even if they used
to accrete from a gaseous fallback disk early on. Another possibility,
of course, is that the effect of the disk on the neutron star is
negligible because it is rapidly disrupted by the pulsar wind. An
important difference between these two evolutionary scenarios for
radio pulsars is that, in the case of disk survival, a remnant disk of
rocks is also expected which could later lead to planet formation, as
explained below.\footnote{Even if the disk survives the pulsar wind,
it is possible for the wind to influence the disk evolution by
providing an extra source of non-thermal ionization.}  We note that,
along those lines, Marsden, Lingenfelter \& Rothschild (2001a; see
also Marsden, Lingenfelter \& Rothschild 2001b) have recently proposed
that torques due to a fallback disk could contribute to spinning down
radio pulsars faster than anticipated from magnetic dipole radiation
alone.

\subsection{Planet Formation}

The cold and metal-rich disk resulting from the evolutionary scenario
outlined in \S3 appears to be an ideal site for the formation of dust,
planetesimals and perhaps planets. The notion that a supernova fallback
disk spreads viscously until it reaches conditions appropriate to the
formation of planets was previously investigated by Lin et
al. (1991). One of the proposed scenarios for the origin of the
planets around the
radio pulsar PSR~1257+12 (Wolszczan \& Frail 1992; Wolszczan 1994)
invokes this same type of fallback disk evolution (see Phinney \&
Hansen 1993 for a review).

The fact that the disk may become neutral and perhaps inviscid when
the thermal ionization instability first sets in affects the planetary
formation scenario dramatically. This is because the disk may not have
time to spread to large enough radii for planets to form. For the
specific initial disk parameters discussed in \S3, the disk outer
radius is $<10^{11}$~cm when it becomes neutral, even for the large
values of the disk initial parameters $R_d(t_0) =10^8$~cm
(corresponding to a specific angular momentum $j =
10^{17}$~cm$^2$~s$^{-1}$ around a solar mass object) and $M_d(t_0)
=10^{-2} M_\odot$. Although the ``strongly-irradiated'' disk model
described in Appendix~A predicts larger final radii, we argued that it
likely overestimates the real effect of irradiation on the disk
evolution.

The tidal disruption radius for a self-gravitating body of density
$\rho \approx 1$~g~cm$^{-3}$ around a neutron star of mass $M_1=1.4
M_\odot$ is (e.g. Aggarwal \& Oberbeck 1974)
\begin{equation}
R_{\rm tid} \approx \left( \frac{M_1}{\rho} \right)^{1/3} \approx
10^{11}~{\rm cm}.
\end{equation}
This shows that, according to our models, a fallback disk will not in
general spread enough during its gaseous phase to allow immediate
planet formation. Hansen (2001; see also Phinney \& Hansen 1993)
recently investigated the viscous evolution of a disk in PSR~1257+12
in scenarios where the disk becomes inviscid below a temperature of
$3000$~K. He considers scenarios which allow more total angular
momentum for the disk than the fallback scenario does and is in 
this way
able to account for the outer planet in PSR~1257+12.  His results are
therefore consistent with ours.

Planet formation will become possible if, later on, the disk of rocks
can spread via collisions beyond the Roche limit. Given that
collisions occur on relatively short, dynamical timescales, one would
naively expect the disk to spread significantly over a system's
lifetime, allowing late planet formation. As long as the disk lies
inside the Roche limit, only small bodies, of the size of rocks or so,
are allowed to form around the young neutron star in the proposed
scenario.\footnote{In that respect, the system of rings around Saturn
may offer some guidance as to the type of behavior to expect. The
collisional dynamics in the deep potential well of a compact object
may be quite different, however.} These objects are held 
together against tidal
disruption by internal material strength. The tidal disruption radius
under these conditions is (e.g. Aggarwal \& Oberbeck 1974)
\begin{equation}
R_{\rm tid, int} \approx \left( \frac{G M_1 \rho r^2}{T} \right)^{1/3}, 
\label{eq:rtidint}
\end{equation}
where $r$ is the rock linear size and T its material strength. For
example, an iron rock of material strength $T \sim
10^{10}$~dyne~cm$^{-2}$ (e.g. Mosenfelder et al. 2000) and size $r
\sim 3 \times 10^6$~cm (roughly corresponding to a mass of
$10^{21}$~g) can exist only beyond $R_{\rm tid, int} \sim
10^{10}$~cm. Only smaller (less massive) rocks are allowed further in,
with a distance-size relation obeying $R_{\rm tid, int} \propto
r^{2/3}$. For example, at a radius of $10^8$~cm, only rocks with $r
\lsim 3 \times 10^3$~cm (mass $\lsim 10^{12}$~g) are allowed.  The
growth of rocks is limited by this maximum allowed mass spectrum.  The
collisional dynamics and global evolution of the disk are likely to be
affected by the Roche-limited growth of its rocks.  We defer a more
detailed discussion of the evolution of such a disk of rocks to future
work.

\subsection{The Black Hole Case}

All the results presented so far were calculated assuming a solar mass
central object, i.e. a neutron star. We discuss here the extension of
our work to the case of a central black hole. The results are
qualitatively similar to the neutron star case.  Assuming a black hole
of mass $M_1=10 M_\odot$, the coefficients $\beta$ and $\beta_{irr}$ in
the global stability criteria given in \S3 become $\approx 1/8$ and
$2/5$, respectively. The amount of fallback during a supernova
explosion resulting in the formation of a black hole is larger than
for a neutron star (see, e.g., Woosley \& Weaver 1995 for examples
with several solar masses of fallback). The numerical simulations of
the evolution of massive stars by Heger, Langer \& Woosley (2000)
suggest that a specific angular momentum $j \sim
10^{17}$~cm$^2$~s$^{-1}$ for the fallback material is also a
reasonable estimate for the black hole case. Below, we give specific
results for a fallback disk around a black hole of mass $M_1=10
M_\odot$, a disk initial mass $M_d(t_0) = 10^{-3} M_\odot$, a disk
initial radius $R_d(t_0) = 10^7$~cm (corresponding to the value of $j$
quoted above) and neglecting the factor of three difference in the
initial viscous timescale (Eq.~\ref{eq:vistime}) for this more massive
central object.

For these parameters, a non-irradiated disk becomes neutral when
\begin{eqnarray}
t_{n} & \approx& 6.3 \times 10^3~{\rm yrs}~(\times T_{c,6}^{-21/37}), \\ 
R_d (t_{n}) & \approx& 1.5 \times 10^{10}~{\rm cm}~(\times T_{c,6}^{6/37}),\\ 
\dot M_d (t_{n}) &\approx& 3.2 \times 10^{16}~{\rm g~s}^{-1}~(\times T_{c,6}^{18/37}),\\
M_d (t_{n}) &\approx& 1.9 \times 10^{-5} \msun ~(\times T_{c,6}^{-3/37}).
\end{eqnarray}

The same disk, but weakly-irradiated, becomes neutral when 
\begin{eqnarray}
t_{n,wi} & \approx& 1.2 \times 10^5~{\rm yrs}~(\times T_{c,6}^{-15/31}), \\ 
R_d (t_{n,wi}) & \approx& 4.5 \times 10^{10}~{\rm cm}~(\times T_{c,6}^{6/31}),\\ 
\dot M_d (t_{n,wi}) &\approx& 8 \times 10^{15}~{\rm g~s}^{-1}~(\times T_{c,6}^{12/31}),\\
M_d (t_{n,wi}) &\approx& 1.5 \times 10^{-5} \msun ~(\times T_{c,6}^{-3/31}).
\end{eqnarray}

We do not consider the strongly-irradiated case: a disk around a black
hole is even less subject to X-ray irradiation than a disk around a
neutron star because there is no central X-ray source but only disk
self-irradiation in this case (Shakura \& Sunyaev 1973). These results
are somewhat different from the neutron star case, but have rather
similar evolutionary implications.

Newly formed black holes surrounded by fallback disks should be active
X-ray sources for relatively short periods of time. The immediate
formation of planets is equally unlikely in the black hole case. Even
for an extreme fallback disk mass of $1 M_\odot$, the maximal radius
reached by the disk is $\sim 5 \times 10^{10}$~cm ($2 \times
10^{11}$~cm) in the non-irradiated (weakly-irradiated) case.  This is
not larger than the tidal disruption radius $R_{\rm tid} \approx 3
\times 10^{11}$~cm for a self-gravitating body of density $\rho
\approx 1$~g~cm$^{-3}$ in that case. Consequently, the outcome of the
evolution of a fallback disk around a black hole is the formation of a
disk of rocks, like in the neutron star case, with a maximum size (or
mass) spectrum increasing with radius. Similarly, the formation of
planets becomes possible only if the disk of rocks spreads beyond the
Roche limit.

The identification of fallback disks around newly-formed black holes
may prove extremely difficult once they become passive. We note,
however, that the collisions experienced by the rocks in the disk will
scatter some of them on orbits such that they will be accreted by the
black hole. Rocks at $10^{11}$~cm with $r \lsim 6 \times 10^7$~cm
(mass $\lsim 10^{25}$~g; see Eq.~[\ref{eq:rtidint}]) would lead to
events of fluence $\lsim 10^{45}$~ergs, while rocks at $10^{8}$~cm ($r
\lsim 2 \times 10^3$~cm; mass $\lsim 3 \times 10^{11}$~g) would lead
to events of fluence only $\lsim 3 \times 10^{31}$~ergs. Of course,
the collisional dynamics of the disk of rocks could limit their growth
and make them, on average, less massive than the maximal value allowed
by tidal forces (assumed here for the fluence estimates). The duration
of the accretion events will depend on the precise orbits of the rocks
as they approach the black hole (see Colgate \& Petschek 1981 and
Tremaine \& Zytkow 1986 for a discussion of comets accreting onto
magnetized neutron stars).

\subsection{Doubly Degenerate Mergers}

The existence of a population of doubly degenerate (white dwarf --
white dwarf) binaries in our galaxy has clearly been established, some
of which are expected to merge within a Hubble time (e.g. Marsh 1995;
Marsh, Dhillon \& Duck 1995). {If the merger does not result in a
Type Ia supernova (Whelan \& Iben 1973; Wheeler 1982; Webbink 1984),
the end product is a massive white dwarf, or a neutron star if
core-collapse occurs.} Numerical simulations show that a large fraction
of the angular momentum of the initial binary ends up in a disk
surrounding the merger product. Whether this disk is composed
primarily of helium or metals, it is subject to the thermal ionization
instability (\S2) and presumably follows the same general evolutionary
scenario as fallback disks (\S3). This is true independent of the
nature of the merged compact object.

The typical mass and angular momentum in a disk resulting from the
merger of two white dwarfs are significantly larger than those
considered for fallback disks in \S3. Benz et al. (1990; see also
Segretain, Chabrier \& Mochkovitch 1997) find that the outer,
rotationally-supported region of the merger product has a mass $\sim
0.3 M_\odot$, at radii $\sim 10^9$~cm. Using these values for the
initial disk mass and radius in the evolutionary model of \S3.1 yields
the following conditions for the disk when it becomes neutral:
\begin{eqnarray}
t_{n} & \approx& 1.8 \times 10^3~{\rm yrs}~(\times T_{c,6}^{-21/37}), \\ 
R_d (t_{n}) & \approx& 3.9 \times 10^{11}~{\rm cm}~(\times T_{c,6}^{6/37}),\\ 
\dot M_d (t_{n}) &\approx& 6 \times 10^{20}~{\rm g~s}^{-1}~(\times T_{c,6}^{18/37}),\\
M_d (t_{n}) &\approx& 1.5 \times 10^{-2} \msun ~(\times T_{c,6}^{-3/37}).
\end{eqnarray}
Consequently, it appears possible for such a disk to spread enough
before becoming neutral to allow planet formation after the activity
period, in regions located beyond the Roche limit at $\sim
10^{11}$~cm. This scenario suggests that massive white dwarfs which
are the result of a doubly degenerate merger are capable of harboring
planets (see Livio, Pringle \& Saffer [1992] for a similar conclusion
but a different line of reasoning). Similarly, {in case of
core-collapse following the merger,} the neutron star end product will
also be a potential host for planets. We note, however, that a
necessary ingredient for planet formation to occur in the doubly
degenerate context is that the secondary in the pre-merger binary was
massive enough to be mostly composed of Carbon/Oxygen rather than
Helium.

\subsection{Other Applications}

In addition to the X-ray point sound discovered recently in Cas A
with the {\it Chandra} satellite, 
Chakrabarty et al. (2000) discuss
three radio-quiet, non-plerionic X-ray
point sources which have been associated with supernova remnants. These
X-ray sources could be explained by emission from a fallback accretion
disk. As noted by Chakrabarty et al., the radio-quiet X-ray point
sources share characteristics in common with AXPs. The absence of
X-ray pulsations may indicate the presence of a black hole accretor in
these systems. We note, however, that the X-ray luminosities of these
four X-ray sources are typically one or two orders of magnitude below
those typical of AXPs. This could be accounted for by the fallback
scenario only for relatively small initial disk mass and angular
momentum.

Fallback disks could also be present around newly-formed compact
objects following core-collapse supernova explosions in binary
systems. We note that, in this context, it is in principle possible
for the fallback material to acquire some extra angular momentum
relative to the compact object during the explosion because of the
non-axisymmetric tidal perturbations induced by the companion. The
presence of the disk will not substantially influence the orbital
evolution of the binary because the angular momentum in the
companion's orbit far exceeds that in the disk. Except perhaps in
close binaries, the fallback disk may not spread enough to interact
strongly with the companion before it becomes neutral (although the
eccentric orbit of the companion, following the supernova explosion,
would favor such an interaction).  Independent of whether the disk is
in the gaseous phase or made of rocks, the effect of the presence of
the companion will be to remove angular momentum from the disk via
tidal interactions (Goldreich \& Tremaine 1980).

\section{Discussion}

We presented in \S3 simple estimates for the age, size and accretion
rates at which supernova fallback disks are expected to become neutral
under the action of the thermal ionization instability. We stress here
that these estimates are only approximate because of the various
simplifying assumptions made in our derivations. We do not know the
exact composition of a fallback disk\footnote{The composition of
the fallback material depends on the amount of mass that makes it in
the disk. For a disk with small mass, say $M\le 10^{-3}M_\odot$, the
fallback material will likely be made of silicon-burning products.},
so we ignore its precise stability properties. The strength of disk
irradiation, which also influences the stability properties, is not
well-determined either. We treated the viscous evolution of an
irradiated disk only in a simplified manner.  We used similarity
solutions to describe the evolution of a disk of finite (and generally
small) radial extent, while the solution applies best to the large
extent case. Although all these approximations are sources of
uncertainty, we find that, quite generally, the model predicts that
fallback disks should become neutral when they are still young,
compact and accreting at generally sub-Eddington rates.

Another {important} source of uncertainty for the models described
in \S3 comes from the early evolution of the fallback disk, when it
accretes at super-Eddington rates. We did not differentiate this phase
from the sub-Eddington one, but the relevance of the similarity
solutions applied to the super-Eddington phase is not guaranteed. A
variety of viscosity laws would predict a power-law evolution for the
accretion rate in the disk with an index close to $-1.2$ as considered
in \S3 (Lynden-Bell \& Pringle 1974), so that this type of solution is
rather robust. On the other hand, the assumption of accretion via a
thin disk may not be valid during the super-Eddington phase. It is
conceivable that the disk evolution will still proceed according to
the power law solutions used in \S3 if the disk finds ways to cool
efficiently, e.g. because of the ``photon bubble'' instability
identified by Gammie (1998), but the evolution during the
super-Eddington phase definitely constitutes an uncertainty for the
models. {If the disk were unable to cool, the initial viscous
evolution would proceed much faster and the global outcome could be
significantly affected.  This uncertainty is essentially captured in
the scaling with $T_{c,6}$ in all our equations.} A rather robust
ingredient of the models appears to be the compact size of fallback
disks when they become neutral. We show explicitly in Appendix~B that
fallback disks with arbitrarily fast power-law evolutions are still
restricted to rather small sizes ($< 1$~AU).

The models described in \S3 assume that the total angular momentum of
the fallback disk is conserved during its evolution. It is possible
for a central neutron star, however, to transfer a fraction of its
spin angular momentum to the disk via magnetospheric interaction. The
angular momentum of the neutron star is
\begin{equation}
A_{\rm NS}=I \Omega \approx 4 \times 10^{47} \left( \frac{P}{15~{\rm
ms}} \right)^{-1} ~{\rm g~cm^2~s^{-1}},
\end{equation}
where $I (\approx 10^{45}~{\rm g~cm^2~s^{-1}})$ and $P$ are the moment
of inertia and spin period of the neutron star, respectively.  The
total angular momentum of the disk is
\begin{eqnarray}
A_{\rm disk} &\approx& M_d(t_0) (G M_{1} R_d(t_0))^{1/2}, \\ &\approx&
3 \times 10^{47}\left( \frac{M_d(t_0)}{10^{-3}~ \msun} \right) \left(
\frac{R_d(t_0)}{10^8~{\rm cm}} \right)^{1/2}~{\rm g~cm^2~s^{-1}},
\end{eqnarray}
where $M_1 = 1.4 M_\odot$ is the mass of the central object. Clearly,
in the case of a rapidly spinning neutron star and a low angular momentum
disk, the contribution from the spun down neutron star to the disk
angular momentum can become important or even dominant over time.

This possibility is not a problem in the magnetospheric interaction
model of Chatterjee et al. (2000) because the initially fast spinning
neutron star loses most of its angular momentum during an early
propeller phase, so that most of the stellar
angular momentum is lost to infinity. By the time the disk
enters the ``tracking phase'' and acquires angular momentum by
spinning down the neutron star,
now having a period of several seconds, the
stellar angular momentum of the neutron star is at most comparable to
that of the disk except for a fallback disk with very small initial
angular momentum.\footnote{The specific AXP disk model discussed in
\S4.1.1 is safe in that respect.} On the other hand, if this early
propeller phase were not to occur, the viscous evolution of a fallback
disk could be mainly determined by the angular momentum that it
receives from the neutron star. In the limit where the disk mass is
conserved and its evolution is driven by the torque applied at its
inner edge by the neutron star, it would follow the power law
solutions found by Pringle (1991) for ``external disks''.

The analogy between fallback disks and disks in close binaries made in
\S3 is also useful if, contrary to what we argued, fallback disks do
not become passive when they become neutral.  If fallback disks were
to possess a residual viscosity in their neutral phase, with a
magnitude significantly smaller than that of the ionized phase
(because, for instance, of a mechanism different from MHD turbulence
operating in the neutral disk), numerical disk instability models
suggest that the disks would then experience outbursts similar to
those of transient close binaries. They would do so until the
accretion rate in the disk is so small that the entire disk is neutral
and no longer subject to the thermal ionization instability. In first
approximation, one would expect the period of evolution with outbursts
to last at least a time comparable to the disk lifetimes estimated in
\S3, although a more detailed numerical investigation is probably
required to answer this question properly. The fact that no such
bright, isolated soft X-ray transient has been discovered to date does
not seem to support this scenario.  On the other hand, if the
magnitude of the viscosity in the neutral phase of fallback disks is
comparable to that in the ionized phase (e.g. because large opacities
leading to efficient MHD coupling), the disks would only experience
very small amplitude luminosity fluctuations due to multiple
reflections of recombination/ionization fronts (Smak 1984; Hameury et
al. 1998; Menou et al. 1999). It may be reasonable to expect the disk
evolution to still proceed in a self-similar manner in that case,
until the central temperature at the disk outer edge reaches a
temperature ($\sim 2000$~K) at which the gas and the magnetic field in
the disk are no longer coupled. This evolutionary scenario cannot be
further characterized without a better knowledge of the opacity in the
neutral metal-rich disk, but we note that it may lead to a passive
neutral disk in a final state not that different from those discussed
in \S3.

We emphasized in \S4.1 an application of the evolutionary models for
fallback disks developed in \S3 to Anomalous X-ray Pulsars
(AXPs). These sources, because of similarities, have been associated
with Soft Gamma Repeaters (SGRs). There is no component in the fallback
accretion disk scenario described here which can explain the burst
phenomenology of SGRs. We also note that Thompson et al. (1999) have
presented strong arguments against standard viscous accretion powering
the quiescent emission of SGRs. It may be possible to explain the SGR
phenomenology by considering the evolution of a fallback disk after it
turns to rocks, but the situation is presently unclear. Along those
lines, we note that Katz, Toole \& Unruh (1994) have proposed an SGR
model in which planetary collisions and subsequent accretion onto a
central neutron star power SGR bursts. We suspect that this scenario
is unlikely to apply to the disk of rocks formed after fallback disk
evolution because of the large mass objects required to power some of
the most luminous SGR bursts observed and the absence of eccentric
orbits favoring large body collisions in the disk (the eccentricities
in the planetary disk predating the supernova explosion in the
scenario described by Katz et al. 1994 are caused by the explosion
itself).

\section{Conclusions}

We have shown that thin accretion disks composed of Carbon, Oxygen and
probably any other metal are subject to the same thermal ionization
instability as Hydrogen and Helium disks. While the instability {is
thought to be} responsible for large-amplitude outbursts in dwarf
novae and X-ray transients, we argued that its role could be different
for supernova fallback disks. As the instability quickly turns such a
metal-rich disk into a neutral medium, there may not be any viscosity
source left in the unmagnetized disk that is sufficient to prevent
dust formation. The outcome of the instability in this context would
therefore be a passive disk of dust which would later evolve into a
disk of rocks via coagulation processes.

We constructed simple analytical models to predict the time at which a
supernova fallback disk becomes neutral. For reasonable values of the
disk initial mass and angular momentum, we find that this happens when
the disks are still young, compact in size and generally accreting at
sub-Eddington rates. A direct application of the models shows that
they can account for disks of age a few $10^3$~yrs, radial extent of a
few $10^9$~cm and accretion rate $\sim 10^{16}$~g~s$^{-1}$, as
inferred for hypothetical disks in Anomalous X-ray Pulsars.

The late evolution of the disk of rocks formed around a young compact
object will strongly depend on its ability to spread via
collisions. The dynamics of this disk may be heavily
influenced by the
fact that it generally lies inside the tidal disruption radius for
self-gravitating bodies (Roche limit), at $\approx 10^{11}$~cm. The
formation of planets around the compact object will become possible
only if/when the disk spreads beyond this Roche limit.

\section*{Acknowledgments} 

We thank Guillaume Dubus, Jeremy Goodman, Brad Hansen, Ramesh Narayan,
Bohdan Paczynski, Eliot Quataert and Scott Tremaine for very useful
discussions.  Support for this work was provided by NASA through
Chandra Fellowship grant PF9-10006 awarded by the Smithsonian
Astrophysical Observatory for NASA under contract NAS8-39073.

\clearpage

\begin{appendix}

\section{Power-Law Evolution of Irradiated Fallback Disks}

In this Appendix, we derive analytical solutions for the evolution of
a supernova fallback disk, under two different assumptions for the
strength of irradiation.

\subsection{Weakly-Irradiated Disk}

In this case, we assume that the disk evolution proceeds as for a
non-irradiated $\alpha-$disk, but the stability criterion including
the heating by irradiation is used (Eq.~[\ref{eq:critirr}]).  A
derivation similar to that presented in \S3.1 shows that the outermost
disk annulus becomes neutral in that case after a time:
\begin{equation}
t_{n,wi} \approx 1.2 \times 10^4~{\rm yrs} ~\left( \frac{M_d
(t_0)}{\beta_{irr} 10^{-3}~\msun} \right)^{16/31} T_{c,6}^{-15/31}
R_{d,8}^{-49/62} (t_0).
\end{equation}
This happens for a disk outer radius
\begin{equation}
R_{d,8} (t_{n,wi}) \approx 1.25 \times 10^3 ~\left( \frac{M_d
(t_0)}{\beta_{irr} 10^{-3}~\msun} \right)^{6/31} T_{c,6}^{6/31}
R_{d,8}^{16/31}(t_0),
\end{equation} 
a disk mass
\begin{equation}
M_d (t_{n,wi}) \approx 2.8 \times 10^{-5}~\msun ~\beta_{irr}^
{3/31}~\left( \frac{M_d (t_0)}{10^{-3}~\msun}
\right)^{28/31} T_{c,6}^{-3/31} R_{d,8}^{15/62} (t_0),
\end{equation} 
and a disk accretion rate
\begin{equation}
\dot M_d (t_{n,wi}) \approx 1.5 \times 10^{17}~{\rm g~s}^
{-1} ~\beta_{irr}^{19/31}~\left( \frac{M_d (t_0)}{ 10^{-3}~\msun}
\right)^{12/31} T_{c,6}^{12/31} R_{d,8}^{32/31} (t_0).
\end{equation}

For a low mass and low angular momentum disk ($M_d (t_0)=10^{-6}~\msun$,
$R_{d,8}(t_0) = 0.01$ and $\beta_{irr} =1$), this gives:
\begin{eqnarray}
t_{n,wi} & \approx& 1.3 \times 10^4~{\rm yrs}~(\times T_{c,6}^{-15/31}), \\ 
R_d (t_{n,wi}) & \approx& 3 \times 10^{9}~{\rm cm}~(\times T_{c,6}^{6/31}),\\ 
\dot M_d (t_{n,wi}) &\approx& 10^{14}~{\rm g~s}^{-1}~(\times T_{c,6}^{12/31}).
\end{eqnarray}

For a high mass and high angular momentum disk ($M_d (t_0)=10^{-2}~\msun$,
$R_{d,8}(t_0) = 1$ and $\beta_{irr} =1$), this gives:
\begin{eqnarray}
t_{n,wi} & \approx& 4 \times 10^4~{\rm yrs}~(\times T_{c,6}^{-15/31}), \\ 
R_d (t_{n,wi}) & \approx& 2 \times 10^{11}~{\rm cm}~(\times T_{c,6}^{6/31}),\\ 
\dot M_d (t_{n,wi}) &\approx& 3.6 \times 10^{17}~{\rm g~s}^{-1}~(\times T_{c,6}^{12/31}).
\end{eqnarray}
As before, it can easily be checked that the disk is still strongly
optically thick to X-rays and its own radiation when it becomes
neutral.

\subsection{Strongly-Irradiated Disk}

We assume now that when the outer radius of the fallback disk reaches
$10^9$~cm, it switches to the more rapid evolution of a vertically
isothermal disk (with a constant irradiation flux to guarantee that
the viscosity is time-independent). Before that, irradiation does not
affect the disk evolution, which proceeds according to
Eqs.~(\ref{eq:evol1}-\ref{eq:evol3}). The properties of the fallback
disk when it reaches $10^9$~cm are therefore given by:
\begin{eqnarray}
t_9 & =& 3.1 \times 10^{-2}~{\rm yrs}~\left[T_{c,6}^{-1} R_{d,8}^{-13/6} (t_0) \right],\\
M_d (t_9)  & = & 0.32  M_d (t_0) \left[ R_{d,8}^{1/2} (t_0) \right],\\
\dot M_d (t_9)  & = & 6.8 \times 10^{23}~{\rm g~s}^{-1} \left[ ~\left(
\frac{M_d (t_0)}{10^{-3}~\msun} \right)
T_{c,6} R_{d,8}^{8/3} (t_0) \right].
\end{eqnarray}

The subsequent viscous evolution of the disk, now assumed to be
vertically isothermal and steadily irradiated, follows:
\begin{eqnarray}
M_d (t)  & = & M_d (t_9) \left( \frac{t}{t_9} \right)^{-7/13}, \label{eq:sievol1}\\
R_d (t)  & = & R_d (t_9) \left( \frac{t}{t_9} \right)^{14/13}, \label{eq:sievol2}\\
\dot M_d (t)  & = & \dot M_d (t_9) \left( \frac{t}{t_9} \right)^{-20/13}.
\label{eq:sissevol} \label{eq:sievol3}
\end{eqnarray}

Using the global stability criterion that includes the heating by
irradiation (Eq.~[\ref{eq:critirr}]), we find that the fallback disk
becomes neutral after a time:
\begin{equation}
t_{n,si} \approx 26.7 ~{\rm yrs} ~\left( \frac{M_d
(t_0)}{\beta_{irr} 10^{-3}~\msun} \right)^{13/48} T_{c,6}^{-35/48}
R_{d,8}^{-13/9} (t_0).
\end{equation}
This happens for a disk outer radius
\begin{equation}
R_{d,8} (t_{n,si}) \approx 1.4 \times 10^4 ~\left( \frac{M_d
(t_0)}{\beta_{irr} 10^{-3}~\msun} \right)^{7/24} T_{c,6}^{7/24}
R_{d,8}^{7/9}(t_0),
\end{equation} 
a disk mass
\begin{equation}
M_d (t_{n,si}) \approx 8.4 \times 10^{-6}~\msun ~\beta_{irr}^{7/48}
\left( \frac{M_d (t_0)}{10^{-3}~\msun} \right)^{41/48} T_{c,6}^{-7/48}
R_{d,8}^{1/9} (t_0),
\end{equation} 
and a disk accretion rate
\begin{equation}
\dot M_d (t_{n,si}) \approx 2 \times 10^{19}~{\rm g~s}^{-1} ~
\beta_{irr}^{5/12} \left( \frac{M_d (t_0)}{10^{-3}~\msun}
\right)^{7/12} T_{c,6}^{7/12} R_{d,8}^{14/9} (t_0).
\end{equation}

For a low mass and low angular momentum disk ($M_d (t_0)=10^{-6}~\msun$,
$R_{d,8}(t_0) = 0.01$ and $\beta_{irr} =1$), this gives:
\begin{eqnarray}
t_{n,si} & \approx& 3.2 \times 10^3~{\rm yrs}~(\times T_{c,6}^{-35/48}), \\ 
R_d (t_{n,si}) & \approx& 5.2 \times 10^{9}~{\rm cm}~(\times T_{c,6}^{7/24}),\\ 
\dot M_d (t_{n,si}) &\approx& 2.7 \times 10^{14}~{\rm g~s}^{-1}~(\times T_{c,6}^{7/12}).
\end{eqnarray}

For a high mass and high angular momentum disk ($M_d (t_0)=10^{-2}~\msun$,
$R_{d,8}(t_0) = 1$ and $\beta_{irr} =1$), this gives:
\begin{eqnarray}
t_{n,si} & \approx& 50 ~{\rm yrs}~(\times T_{c,6}^{-35/48}), \\ 
R_d (t_{n,si}) & \approx& 2.7 \times 10^{12}~{\rm cm}~(\times T_{c,6}^{7/24}),\\ 
\dot M_d (t_{n,si}) &\approx& 7.7 \times 10^{19}~{\rm g~s}^{-1}~(\times T_{c,6}^{7/12}).
\end{eqnarray}
Again, the disk is strongly optically thick to X-rays and its own
radiation when this happens.

\clearpage

\section{Disk Radial Extent for an Arbitrarily Fast Power-Law Evolution}

In this Appendix, we show that the radial extent of a fallback disk,
when it becomes neutral, is finite and rather small even for an
arbitrarily fast power law evolution in time. The derivation is made
using the stability criterion for a non-irradiated disk
(Eq.~[\ref{eq:crit}]) but a comparable result would be obtained using
the criterion for an irradiated disk.

We parameterize the disk viscous evolution as follows:
\begin{eqnarray}
M_d (t)  & = & M_d (t_0) \left( \frac{t}{t_o} \right)^{-p}, \label{eq:sievol11}\\
R_d (t)  & = & R_d (t_0) \left( \frac{t}{t_0} \right)^{2p}, \label{eq:sievol22}\\
\dot M_d (t)  & = & \dot M_d (t_0) \left( \frac{t}{t_0} \right)^{-(1+p)},
\label{eq:sissevol2} \label{eq:sievol33}
\end{eqnarray}
where $p>0$ for convergence. The time $t_n$ at which the disk becomes
neutral is given by (Eq.~[\ref{eq:dead}]):
\begin{equation}
\frac{t_n}{t_0} \approx \left( R_{d,10}^3 (t_0) \frac{\beta
10^{16}~{\rm g~s}^{-1}}{\dot M_d (t_0)} \right) ^{\frac{-1}{1+7p}}.
\end{equation}

The radius at which the outer disk becomes neutral is
\begin{equation}
R_d(t_n) \approx R_d(t_0) \left( R_{d,10}^3 (t_0) \frac{\beta
10^{16}~{\rm g~s}^{-1}}{\dot M_d (t_0)} \right)^{\frac{-2p}{1+7p}},
\end{equation}
with an asymptotic power law index of $-2/7$ at large $p$. In this
limit, the corresponding value for a low mass and angular momentum
disk ($M_d (t_0)=10^{-6}~\msun$, $R_{d,8}(t_0) = 0.01$) is $R_d(t_n)
\approx 5 \times 10^9$~cm, while it yields $R_d(t_n) \approx
10^{12}$~cm for a disk with high mass and angular momentum ($M_d
(t_0)=10^{-2}~\msun$, $R_{d,8}(t_0) = 1$). This shows that disks with
power-law evolutions faster than assumed in \S3 are still restricted
to {\bf sub-AU} sizes.

\end{appendix}

\clearpage

\begin{thebibliography}{}
\bibitem[]{}Aggarwal, H.R. \& Oberbeck, V.R. 1974, ApJ, 191, 577
\bibitem[]{}Alexander, D.R. \& Ferguson, J.W. 1994, ApJ, 437, 879
\bibitem[]{}Alpar, M.A. 1999, preprint, astro-ph/9912228 
\bibitem[]{}Alpar, M.A. 2000, ApJ, submitted, astro-ph/0005211
\bibitem[Balbus \& Hawley (1991)]{bh91}Balbus, S. A., \& Hawley, J.F. 1991, \apj, 376, 214
\bibitem[Balbus \& Hawley (1997)]{bh97}Balbus, S. A., \& Hawley, J.F. 1998, Rev. Mod. Phys., 70, 1
\bibitem[]{}Balbus, S. A. \& Terquem, C. 2001, ApJ, submitted, astro-ph/0010229
\bibitem[]{}Benz, W., Bowers, R.L., Cameron, A.G.W. \& Press, W.H. 1990, ApJ, 348, 647
\bibitem[]{}Cabot, W. 1996, ApJ, 465, 874
\bibitem[]{}Cannizzo, J.K. 1984, Nature, 311, 443
\bibitem[]{}Cannizzo, J.K. 1993a, ApJ, 419, 318
\bibitem[]{}Cannizzo, J.K. 1993b, in Wheeler J.C., ed., Accretion discs in Compact Stellar Systems (World Scientific, Singapore), p.~6
\bibitem[]{}Cannizzo, J.K., Lee, H.M. \& Goodman, J. 1990, ApJ, 351, 38
\bibitem[]{}Chakrabarty, D., Pivovaroff, M.J., Hernquist, L.E., Heyl, J.S. \& Narayan, R., 2001, ApJ, in press, astro-ph/0001026
\bibitem[]{}Chatterjee, P. \& Hernquist, L. 2000, ApJ, 543, 368
\bibitem[]{}Chatterjee, P., Hernquist, L. \& Narayan, R. 2000, ApJ 534, 373
\bibitem[]{}Chevalier, R.A. 1989, ApJ, 346, 847
\bibitem[]{}Colgate, S.A. 1988, in {\it Supernova 1987A in the Large Magellanic Cloud; Proceedings of the Fourth George Mason Astrophysics Workshop, Fairfax}, p.~341
\bibitem[]{}Colgate, S.A. \& Petschek, A.G. 1981, ApJ, 248, 771
\bibitem[]{}Cunningham, C. 1976, ApJ, 208, 534 
\bibitem[]{}Dubus, G., Hameury, J.-M. \& Lasota, J.-P., 2001, preprint
\bibitem[]{}Dubus, G., Lasota, J.-P., Hameury, J.-M. \& Charles, P. 1999, MNRAS, 303, 139
\bibitem[]{}El-Khoury, W. \& Wickramasinghe, D. 2000, A\&A, 358, 154
\bibitem[]{}Fleming, T.P., Stone, J.M. \& Hawley, J.F., 2000, ApJ, 530, 464
\bibitem[]{}Frank, J., King, A. \& Raine, D., 1992, Accretion Power in Astrophysics (Cambridge Univ. Press, Cambridge)
\bibitem[]{}Fryer, C.L. \& Heger, A. 2000, 541, 1033
\bibitem[Gammie (1996)]{gam96}Gammie, C.F. 1996, \apj, 457, 355
\bibitem[]{}Gammie, C.F. 1998, MNRAS, 297, 929
\bibitem[]{}Gammie C.F. \& Menou K., 1998, ApJ, 492, L75
\bibitem[]{}George, I.M. \& Fabian, A.C. 1991, MNRAS, 249, 352
\bibitem[]{}Goldreich, P. \& Tremaine, S. 1980, ApJ, 241, 425
\bibitem[]{}Goodman, J. 1993, ApJ, 406, 596
\bibitem[]{}Hameury J.-M., Menou K., Dubus G., Lasota J.-P., Hur\'e J.-M., 1998, MNRAS, 298, 1048
\bibitem[]{}Hansen, B.M.S. 2001, in {\it Stellar Collisions, Mergers and their Consequences, ASP Conference Series}, ed. M. Shara, astro-ph/0008226 
\bibitem[]{}Hawley, J.F., Balbus, S.A. \& Winters, W.F. 1999, ApJ, 518, 394 
\bibitem[Hawley, Gammie, \& Balbus (1996)]{hgb96} Hawley, J. F., Gammie, C. F., \& Balbus, S. 1996, \apj, 464, 690
\bibitem[]{}Heger, A., Langer, N. \& Woosley, S.E. 2000, ApJ, 528, 368
\bibitem[]{}Heyl, J.S. \& Hernquist, L. 1997, ApJ, 489, L67
\bibitem[]{}Hulleman, F., van Kerkwijk, M.H. \& Kulkarni, S.R. 2000b, Nature, 408, 689
\bibitem[]{}Hulleman, F., van Kerkwijk, M.H., Verbunt, F.W.M. \& Kulkarni, S.R. 2000a, A\&A, 358, 605
\bibitem[]{}Iglesias, C.A. \& Rogers, F.J. 1996, ApJ, 464, 943
\bibitem[]{}Illarionov, A.F. \& Sunyaev, R.A. 1975, A\&A, 39, 185
\bibitem[]{}Katz, J.I., Toole, H.A. \& Unruh, S.H. 1994, ApJ, 437, 727
\bibitem[]{}Lasota, J.-P. 2001, New Astron. Rev., in press, astro-ph/0102072
\bibitem[]{}Lin, D.N.C., Woosley, S.E. \& Bodenheimer, P.H. 1991, Nature, 353, 827
\bibitem[]{}Livio, M., Pringle, J.E. \& Saffer, R.A. 1992, MNRAS, 257, 15
\bibitem[]{}Ludwig, K., Meyer-Hofmeister, E. \& Ritter, H. 1994, A\&A, 290, 473
\bibitem[]{}Lynden-Bell, D. \& Pringle, J.E., 1974, MNRAS, 168, 603
\bibitem[]{}Marsden, D., Lingenfelter, R.E. \& Rothschild, R.E. 2001a, ApJL, in press, astro-ph/0008300
\bibitem[]{}Marsden, D., Lingenfelter, R.E. \& Rothschild, R.E. 2001b, preprint, astro-ph/0102049
\bibitem[]{}Marsh, T.R., 1995, MNRAS, 275, L1
\bibitem[]{}Marsh, T.R., Dhillon, V.S. \& Duck, S.R., 1995, MNRAS, 275,828
\bibitem[]{}Menou, K., 2000, Science, 288, 2022
\bibitem[]{}Menou K., Hameury J.-M., Stehle R., 1999, MNRAS, 305, 79
\bibitem[]{}Menou, K. \& Quataert, E. 2001, ApJ, in press, astro-ph/0008368
\bibitem[]{}Meyer F., Meyer-Hofmeister E., 1981, A\&A, 104, L10
\bibitem[]{}Meyer-Hofmeister E., 1992, A\&A, 253, 459
\bibitem[]{}Michel, F.C. 1988, Nature, 333, 644
\bibitem[]{}Michel, F.C. \& Dessler, A.J. 1981, ApJ, 251, 654
\bibitem[]{}Michel, F.C. \& Dessler, A.J. 1983, Nature, 303, 48
\bibitem[]{}Mihalas, D., Hummer, D.G., Mihalas, B.W. \& Dappen, W. 1990, ApJ, 350,300
\bibitem[]{}Mineshige, S., Nomoto, K. \& Shigeyama, T., 1993, A\&A, 267, 95
\bibitem[]{}Mosenfelder, J.L., Connolly, J.A.D., Rubie, D.C. \& Liu, M. 2000, Phys. Earth Plan. Interiors, 120, 63
\bibitem[]{}Narayan, R., Mahadevan, R. \& Quataert, E., 1998b, in The Theory of Black Hole Accretion Discs, eds. M. A. Abramowicz, G. Bjornsson, and J. E. Pringle (Cambridge: Cambridge University Press), astro-ph/9803141.  
\bibitem[]{}Osaki, Y. 1996, PASP, 108, 39
\bibitem[]{}Perna, R., Hernquist, L., \& Narayan, R. 2000, ApJ, 541, 344
\bibitem[]{}Perna, R. \& Hernquist, L. 2000, ApJ, 544, L57
\bibitem[]{}Phinney, E.S. \& Hansen, B.M.S. 1993, in {Planets around pulsars, Proceedings of the Conference, Caltech, Pasadena}, p. 371-390.
\bibitem[]{}Pringle, J.E. 1974, Ph.D. Thesis, University of Cambridge 
\bibitem[]{}Pringle, J.E. 1991, MNRAS, 248, 754
\bibitem[]{}Richard, D. \& Zahn, J.-P. 1999, A\&A, 347, 734
\bibitem[]{}Ryu, D. \& Goodman, J. 1992, ApJ, 388, 438
\bibitem[]{}Segretain, L., Chabrier, G. \& Mochkovitch, R. 1997, ApJ, 481, 355
\bibitem[]{}Shakura, N.I. \& Sunyaev, R.A. 1973, A\&A, 24, 337
\bibitem[]{}Smak J., 1983, Acta Astron., 33, 333
\bibitem[]{}Smak J., 1984, Acta Astron., 34, 161
\bibitem[Spruit (1987)]{spr87} Spruit, H C. 1987, A\&A, 184, 173
\bibitem[]{}Stone, J.M. \& Balbus, S.A. 1996, ApJ, 464, 364
\bibitem[]{}Thompson, C. \& Duncan, R.C. 1996, ApJ, 473, 322
\bibitem[]{}Thompson, C. et al. 2000, ApJ, 543, 340
\bibitem[]{}Tremaine, S. \& Zytkow, A.N. 1986, ApJ, 301, 155
\bibitem[]{}Tsugawa, M. \& Osaki, Y. 1997, PASJ 49, 75
\bibitem[]{}van Paradijs, J. 1996, ApJL, 464, L139
\bibitem[]{}Vrtilek, S.D., Raymond, J.C., Garcia, M.R., Verbunt, F., Hasinger, G. \& Kurster, M. 1990, A\&A, 235, 162
\bibitem[]{}Warner, B. 1995, {\it Cataclysmic variable stars} (Cambridge University Press, Cambridge)
\bibitem[]{}Warner, B., Nather, R. E., \& MacFarlane, M. 1969, Nature, 222, 233
\bibitem[]{}Webbink, R.F., 1984, ApJ, 277, 355
\bibitem[]{}Wheeler, J.C., 1982, in {\it Supernovae: A survey of current research}, eds. M.J. Rees \& R.J. Stoneham (Dordrecht: Reidel), p. 167
\bibitem[]{}Whelan, J. \& Iben, I.Jr., 1973, ApJ, 186, 1007
\bibitem[]{}Wolszczan, A. 1994, Science, 264, 538
\bibitem[]{}Wolszczan, A. \& Frail, D.A. 1992, Nature, 355, 145
\bibitem[]{}Woosley, S.E. \& Weaver, T.A. 1995, ApJS, 101, 181
\end{thebibliography}
\end{document}